\RequirePackage{fix-cm}
\documentclass[conference]{IEEEtran}
\IEEEoverridecommandlockouts
\usepackage{color}
\usepackage{bbding}
\usepackage{graphicx}
\usepackage{amsmath}
\usepackage{amssymb}
\usepackage{float}
\usepackage{stfloats}
\usepackage{algorithm}
\usepackage{textcomp}
\usepackage{algpseudocodex}
\usepackage{cite} 
\usepackage{soul}
\usepackage{upgreek}
\usepackage{balance}
\usepackage{hyperref}

\hypersetup{
    pdftitle={Inferring Ingrained Remote Information in AC Power Flows Using Neuromorphic Modality Regime},    
    pdfauthor={},     
    pdfsubject={},   
    pdfcreator={},   
    pdfproducer={},  
    pdfkeywords={}, 
}

\IEEEaftertitletext{\vspace{-2\baselineskip}}

\def\BibTeX{{\rm B\kern-.05em{\sc i\kern-.025em b}\kern-.08em
    T\kern-.1667em\lower.7ex\hbox{E}\kern-.125emX}}
\begin{document}

\title{Inferring Ingrained Remote Information in AC Power Flows Using Neuromorphic Modality Regime\\}

\author{\IEEEauthorblockN{Xiaoguang Diao, Yubo Song and Subham Sahoo}
\IEEEauthorblockA{\textit{Department of Energy, Aalborg University, Denmark} \\
e-mail: 2018202070081@whu.edu.cn, \{\texttt{yuboso,sssa}\}@energy.aau.dk}\\
}

\maketitle

\begin{abstract}
In this paper, we infer remote measurements such as remote voltages and currents online with change in AC power flows using spiking neural network (SNN) as grid-edge technology for efficient coordination of power electronic converters. This work unifies power and information as a means of data normalization using a multi-modal regime in the form of spikes using energy-efficient neuromorphic learning and event-driven asynchronous data collection. Firstly, we organize the synchronous real-valued measurements at each edge and translate them into asynchronous spike-based events to collect sparse data for training of SNN at each edge. Instead of relying on error-dependent supervised data-driven learning theory, we exploit the latency-driven unsupervised Hebbian learning rule to obtain modulation pulses for switching of power electronic converters that can now comprehend grid disturbances locally and adapt their operation without requiring explicit infrastructure for global coordination. Not only does this philosophy block exogenous path arrival for cyber attackers by dismissing the cyber layer, it also entails converter adaptation to system reconfiguration and parameter mismatch issues. We conclude this work by validating its energy-efficient and effective online learning performance under various scenarios in different system sizes, including modified IEEE 14-bus system and under experimental conditions.

\end{abstract}


\IEEEpeerreviewmaketitle

\section{Introduction}

\IEEEPARstart{D}{igitalization} of power electronic dominated grids has brought new data-driven directions in health monitoring and coordination of ever-scaling distributed energy resources (DERs). As we move towards decentralized generation, the control and reliability of DERs demand orchestration from communication networks as an overlaying control and coordination layer to cope with the ever-increasing power demand. However, these digitization efforts not only have increased cyber attack vulnerabilities but have added more unreliability performance in the form of latency, data dropouts, etc {\cite{sahoo}, \cite{cybbook}}. 

To this end, many cyber security detection and mitigation technologies for power electronic dominated grids have surfaced that can be augmented easily into the cyber \cite{ye} as well as the physical layer \cite{sahoo2022cyber}. However, the pace of evolving cyber-physical vulnerabilities can outgrow these research developments quite easily. Apart from security concerns, information exchange enabled by the cyber layer also faces reliability concerns with communication network traffic leading to latency, data dropouts, link failure, etc. As a result, the stability and reliability of power electronic dominated grids is affected \cite{sahoo}, \cite{yubo}. Although decentralized coordination using droop control strategy \cite{7500071} addresses the said concerns in a monolithic manner, its error-immanent operation affects the system performance and efficiency. This mandates an intrinsic communication principle that can effectively unify power \& information to improve power electronic system cognition and resiliency. 

Since the layered hierarchy of communication networks overlaying power grids introduces complexity in modeling, \textit{Power Talk} and \textit{Talkative Power} technologies were developed for converters to communicate via tie-lines such that the information is tactfully modulated over the power signals \cite{PowerTalk}. The communication rate is improved by modifying the switching characteristics. Several innovations involving the co-transfer of power \& data via transmission lines\footnote{They are commonly referred to as \textit{tie-lines}. Both terms are interchangeably used in this paper as per the context.} are inspired by this technology can be found in \cite{marco}. However, the scalability of these co-transfer technologies is still an open question.

To address the scalability problem, we take inspiration from biologically-plausible neuromorphic architecture as an extension of our previous work \cite{sahoo2024nsc}, \cite{sahoo2024spike} for DC systems, and use spiking neural network (SNN) at each edge for multi-modal information exchange between DERs via tie-lines in an AC system and achieve coordinated control. The multi-modal inference scheme is steered by some of the key event-driven \textit{semantic} metrics proposed in \cite{kirti1}, \cite{kirti2}.

\begin{figure*}[!thb]\centering
	\includegraphics[width=0.85\linewidth]{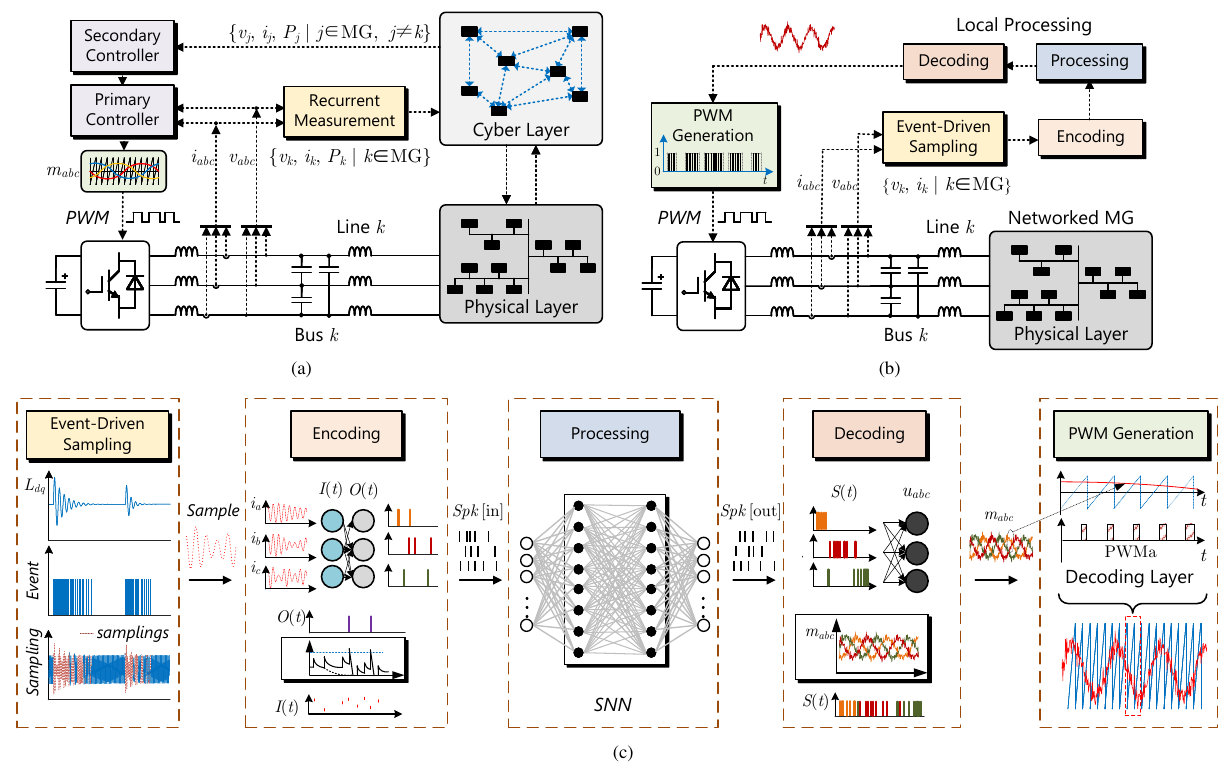}
    \caption{Structural elements of the proposed modality of power and ingrained remote information: (a) conventional cyber-physical control framework, (b) simplified architecture due to the proposed co-transfer methodology, (c) design stages from sampling to PWM generation using Hebbian learning based weight update policy in SNN at each edge.}
        \label{fig_1}
\end{figure*}

Going beyond the conventional cyber-physical hierarchical control for power electronic dominated grids in Fig. 1(a), we convert dynamic grid disturbances into binary and sparse \textit{asynchronous events} \cite{EventDriven} using local measurements, that are then introduced as \textit{spikes} based data for online training and adapting the SNNs deployed at each DER. Locally generated spikes customize SNN at each edge in an online fashion using unsupervised Hebbian learning rule that comprehends the power grid network disturbances locally, and adapts DERs' response. Furthermore, the adaptation features allow it to be a promising energy-efficient grid-edge intelligence tool that can be integrated in a plug-and-play manner. In summary, the proposed technology offers the following features:

\begin{enumerate}
    \item Dynamic remote estimation with energy-efficient neuromorphic event-driven grid-edge processor deployed at each DER. This is performed by inferring ingrained remote information from the AC power flow itself at each edge. To train the SNN, the real-valued measurements are converted into asynchronous spikes for data normalization using a modality of power and information embedded into the timing notion of a spike occurrence. We exploit unsupervised Hebbian learning based rules to update the weights of SNN at each edge. As a result, SNNs can be incrementally deployed at each node as a decentralized secondary controller that significantly reduces hierarchical control stages (see Fig. 1(b)). By completely dismissing the cyber layer for system operation, it bypasses the cyber unreliability concerns and attack vulnerabilities altogether.
    \item Since it replaces the conventional cyber-physical hierarchical stages for control, it also reinforces adaptive behavior into each DER to reconfigure its operation during scenarios, such as system reconfiguration and parameter mismatch issues.
\end{enumerate}

\section{Design Preliminaries of Multi-Modal Power in AC Systems}

As shown in Fig. \ref{fig_1}(a), in comparison with the conventional cyber-physical control framework for power electronic dominated grids, ingrained remote information in AC power flow can instill a very simplified architecture for each DER, which has been shown in Fig. \ref{fig_1}(b). However, we need grid-edge tools that can effectively comprehend remote information, such as remote voltages, currents, power and exploit it for coordinated control of power electronic dominated grids. This is where we deploy a spiking neural network (SNN) at each edge so that we can instill an event-based selective data collection. More specifically, the SNNs at each edge will update their weights in accordance with dynamic conditions in the system. Although the dynamic change is global, the impact on DERs might be local, which then condenses it to local governing equations for generation of spikes at each DER. Since it equips a spike-to-spike based SNN control strategy for each DER, it overrules many hierarchical stages from the conventional cyber-physical control framework in Fig. \ref{fig_1}(a). As shown in Fig. \ref{fig_1}(c), we exploit an event-driven sampling policy to selectively capture the most significant data corresponding to dynamic disturbances only. These analog voltage and current data are then transformed into spikes through the encoding layer, as shown in Fig. \ref{fig_1}(c). For example, it can be seen in Fig. \ref{fig_1}(c) that spikes are only generated as per the \textit{event-driven sampling criteria} only when a dynamic change is encountered in the \textit{dq} signals.

\begin{figure}[!t]
\centering
	\includegraphics[width=0.85\linewidth]{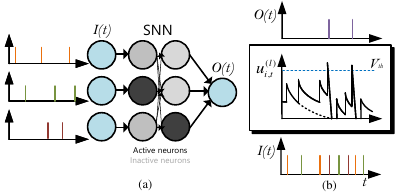}
	\caption{Leaky integrate and fire (LIF) model of SNN: (a) Input and output of the SNN at each DER, where only some LIF neurons are active subject to the frequency of spikes, (b) the neuron model that synthesizes input spikes into selective output spikes based on a voltage threshold $\mathrm{V_{th}}$.}
    \label{fig_2}
\end{figure}

The learning activation in SNN is performed by the biologically plausible leaky integrate-and-fire (LIF) neuron model that learns by performing spatio-temporal integration of the synaptic inputs, as shown in Fig. \ref{fig_2}. Approximating the LIF neuron dynamics using the Euler method, we get:
\begin{equation}
    \mathrm{\tau_m \frac{\mathrm{d}V_{mem}}{\mathrm{d}t} = -(V_{mem} - V_{th}) + \frac{I_s}{g_L}}
\end{equation}
where, $\mathrm{V_{mem}}$ is the membrane potential, $\mathrm{g_L}$ is the leaky conductance of each membrane, $\mathrm{V_{th}}$ is the synaptic threshold, $\mathrm{I_s}$ is the synaptic current and $\mathrm{\tau_m}$ is the membrane time constant. Similar to a switched \textit{RC} network with the same voltage potential, each neuron can be programmed with a certain stored charge $\mathrm{q}$ that has a corresponding notion of leakage, which only gets activated beyond a given voltage threshold $\mathrm{V_{th}}$. The equivalence with (1) for a switched \textit{RC} network can be given by:
\begin{equation}
    \mathrm{\tau_{RC} \frac{\mathrm{d}q}{\mathrm{d}t} = -\frac{V_{mem} - V_{th}}{C} + {q}}
\end{equation}
where, $\mathrm{\tau_{RC} = RC}$. As it can be seen in Fig. \ref{fig_2}, an output spike is generated only when the threshold voltage $\mathrm{V_{th}}$ is reached. Hence, the SNN becomes a sparse and energy-efficient alternative, owing to the fact the LIF neurons are active only when: (a) input events are present, (b) culmination of the input events goes beyond the activation potential $\mathrm{V_{th}}$.

This approach enables the DER to dynamically adapt its operation in response to various disturbances. It is worth notifying that during steady-state conditions, both event-driven sampling in the input and the SNN remain idle, consuming minimal energy. 

\subsection{Event-Driven Sampling}

To carry out a structured and sparse mechanism for data collection, we formalize only the most significant data to be triggered as binary events, that correspond to any physical disturbance in this paper. For instance, the event $\mathrm{\Omega_{vd}}$, associated with the $\mathrm{d}$-axis component $\mathrm{v_d}$ of sampled voltage (see Algorithm \ref{alg_eventcap}), can be detected using a moving window approach using:
\begin{equation}
    \mathrm{\Omega_{vd} : \frac{1}{N_w}\sum{(\mathbf{v_{dw}}-\mathbf{\Bar{v}_{dw}})^2} > \sigma^{V_{d}}_{th}}
\end{equation}
where, $\mathrm{\mathrm{N_w}}$ is the length of the moving window.

\textit{\textbf{Remark:} It is worthy notifying that a moving window approach for event detection and spike generation has been adopted in this work to achieve a good balance between implementation efficiency, versatility and biophysical accuracy of the neurons \cite{loihi}. The data movement rather assists the solver-based digital implementation tasks.}

\begin{algorithm}[!t]
\caption{Event-driven SNN activation for Inverter $\mathrm{k}$}
\label{alg_eventcap}
\begin{algorithmic}[1]
\Require $\mathrm{m^\mathrm{th}}$ samples of system state (\textit{dq} voltage \& current) $\mathrm{\mathbf{x} = [\mathbf{v_{dq}}\;\mathbf{i_{dq}}]^\mathbf{T}}$,  length of the moving window $\mathrm{N_w}$, and thresholds of voltage \& current variance to trigger an event $\mathrm{\sigma^{V_{dq}}_{th}}$, $\mathrm{\sigma^{I_{dq}}_{th}}$.
\State $\mathrm{\mathbf{x_{w}} \gets \{\mathbf{x}[m-N_w],...,\mathbf{x}[m-1]\}}$, $\mathrm{\mathbf{\Bar{x}_{w}}=\frac{1}{N_w}\sum{\mathbf{x_{w}}}}$ 
\State Verify $\mathrm{\Omega_{v_d}}$, $\mathrm{\Omega_{v_q}}$, $\mathrm{\Omega_{i_d}}$, $\mathrm{\Omega_{i_q}}$ and $\mathrm{\Omega}$ by Eqs. (1) and (2)
\If {$\mathrm{\Omega}$}
    \State \texttt{event} starts
    \Repeat
        \State Activate SNN
        \State Transfer $\mathrm{\mathbf{x}_{k}[m]}$ to SNN
        \State Update $\mathrm{\mathbf{x_{w}}}$ and $\mathrm{\mathbf{\Bar{x}_{w}}}$ by the $\mathrm{m}$-th sample:
        \State $\mathrm{\mathbf{x_{w}} \gets \{\mathbf{x}[m-N_w+1],...,\mathbf{x}[m]\}}$
        \State $\mathrm{\mathbf{\Bar{x}_{w}}=\frac{1}{N_w}\sum{\mathbf{x_{w}}}}$
        \State $\mathrm{m \gets m+1}$
        \State Re-verify {$\mathrm{\Omega}$} by Eqs. (1) and (2)
    \Until {$\mathrm{\texttt{NOT}\:\Omega}$}
    \State Deactivate SNN
    \State \texttt{event} ends
\EndIf
\end{algorithmic}
\end{algorithm}

When a dynamic change occurs in $\mathrm{v_d}$, $\mathrm{\Omega_{vd}}$ is triggered to a transition from \texttt{0} to \texttt{1}, and returns to \texttt{0} when $\mathrm{v_d}$ reaches a new steady state. A similar approach is applied to detect dynamics in $\mathrm{v_q}$, $\mathrm{i_d}$, and $\mathrm{i_q}$, denoted as $\mathrm{\Omega_{vq}}$, $\mathrm{\Omega_{id}}$, and $\mathrm{\Omega_{iq}}$, respectively. Event-driven sampling \cite{kirti1} and sparse data collection is subsequently triggered by $\mathrm{\Omega}$, which can be given by:
\begin{equation}
     \mathrm{\Omega =\Omega_{vd}  \quad  \texttt{OR} \quad  \Omega_{vq}  \quad \texttt{OR} \quad \Omega_{id} \quad  \texttt{OR} \quad  \Omega_{iq}}
     \label{eq_event}
\end{equation}

Using (\ref{eq_event}), only when $\mathrm{\Omega}$ is \texttt{1}, the corresponding dynamics in the measurements $\mathrm{\{v_k,i_k \ \forall \ k \in N_\mathrm{DER}\}}$ of inverter $\mathrm{k}$ will be sampled for inputs to the inferential learning of SNN, as described in {Algorithm \ref{alg_eventcap}}. 

\subsection{Encoding, Decoding, and Training of SNN}

We utilize the spiking response model (SRM) \cite{skatchkovsky2021spiking} in this paper, which is a widely recognized neuron model that represents the complex and biologically plausible neurons.

\subsubsection{Encoding of Input Samplings}
As depicted in Fig. \ref{fig_1}(c), we employ rate encoding method to convert the real-valued samples $\mathrm{v_k}$ and $\mathrm{i_k}$ into spikes. In rate encoding, a higher input value results in a greater number of spikes within a specific time interval. For instance, considering $\mathrm{i_a}$ as an example in Fig. \ref{fig_1}(c), spikes are generated only when $\mathrm{i_a[m]}$ exceeds the carrier wave.
 
\subsubsection{Training of SNN}
In Fig. \ref{fig_1}(c), the input to SNN comprises spikes originating from the encoding layer, which encapsulates crucial dynamic information. The final objective behind the training process is to shape these spikes into a PWM signal. To achieve this, we utilize modulation signals $\mathrm{\mathbf{m_{abc}}}$ in Fig. \ref{fig_1}(a) as the regression target to initiate SNN, obtained from the control of virtual synchronous generator (VSG).  Since $\mathrm{\mathbf{m_{abc}}}$ is an analog signal, we employ the membrane potential $\mathrm{\mathbf{u_{abc}}}$ (see Fig. 1(c)) of three neurons in the output layer to align with it. Training the SNN involves the well-known back-propagation algorithm in conjunction with the mean squared error (MSE) loss function:
\begin{equation}
    \mathrm{\texttt{Loss}=\frac{1}{N_w}\sum_{i=1}^{N}(\mathbf{m_{abc}}-\mathbf{u_{abc}})^2}
    \label{Eq_5}
\end{equation}

\subsubsection{Decoding of SNN}
To convert the membrane potential $\mathrm{\mathbf{u_{abc}}}$ into spikes for PWM generation, we employ a dedicated threshold signal, designed to match the shape of the carrier wave as shown in Fig. \ref{fig_1} using:

\small
\begin{equation}
    \mathrm{\texttt{Threshold} = \cfrac{2}{T_s}\left[t-\cfrac{2N-1}{2}T_s\right]},\; \mathrm{(N-1)T_s \leq t<NT_s}
    \label{Eq_6}
\end{equation}
\normalsize
where, $\mathrm{T_s}$ represents the carrier signals' period. We adapt the membrane potential model from \cite{skatchkovsky2021spiking} by eliminating the influence of output spikes. When the membrane potential surpasses the $\mathrm{\texttt{Threshold}}$ in (6), spikes are generated. This neuron model aligns with the modulation process of the sinusoidal pulse width modulation (SPWM). Thus, if the membrane potential accurately matches $\mathrm{\mathbf{m_{abc}}}$, the output spikes will also map the PWM signals accordingly.

\section{Performance Validation}

\begin{figure}[!t]
    \centering
	\includegraphics[width=\linewidth]{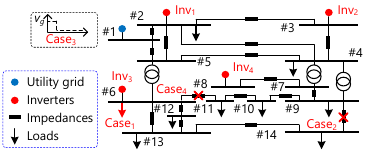}
	\caption{Test cases for various simulation scenarios in Section III carried out on DER at bus \#6 in modified IEEE 14-bus system.}
    \label{fig_3}
\end{figure}

The proposed methodology is validated through a co-simulation of Python and Matlab. The SNN model is built using Python and then integrated into the Simulink environment for simulations. The state of the inverter at Bus \#6 in the IEEE 14-bus system in Fig. \ref{fig_3} is used for verification. With reactive power sharing and average voltage regulation being the coordination objectives of the secondary controller for each inverter, fault ride-through (FRT) capability \cite{liu2021current} is also seamlessly integrated into the SNN.

The system parameters of IEEE 14-bus system can be referred to from {\cite{para}}. The deployed SNN at each DER is designed as: \{$\mathrm{\mathbf{Layer}_\mathrm{in,\;6\times 1}}$, $\mathrm{\mathbf{Layer}_\mathrm{hidden,\;I \& II,\;256\times 2}}$, $\mathrm{\mathbf{Layer}_\mathrm{out,\;3\times 1}}$\}. The dimension of input data is $\mathrm{\mathbf{D}_\mathrm{in,\;15000\times 6}}$ = \{$\mathrm{\mathbf{v}_{abc,\;15000\times 3}}$, $\mathrm{\mathbf{i}_{abc,\;15000\times 3}}$\} including 15000 set-points and 6 variables. The dimension of output data is $\mathrm{\mathbb{D}_\mathrm{out,\;15000\times 3}}$ = \{$\mathrm{\mathbf{m}_{a,\;15000\times 1}}$, $\mathrm{\mathbf{m}_{b,\;15000\times 1}}$, $\mathrm{\mathbf{m}_{c,\;15000\times 1}}$\}. The length of the moving window $\mathrm{N_w}$ in Algorithm \ref{alg_eventcap} is 4000 samples with a sampling frequency of 100 kHz. Furthermore, the thresholds used in Algorithm \ref{alg_eventcap} are $\mathrm{\sigma^{V}_{th}=0.15}$, $\mathrm{\sigma^{I}_{th}=0.05}$.

\subsection{Coordination Control Results}

\begin{figure}[!t]
    \centering
	\includegraphics[width=0.85\linewidth]{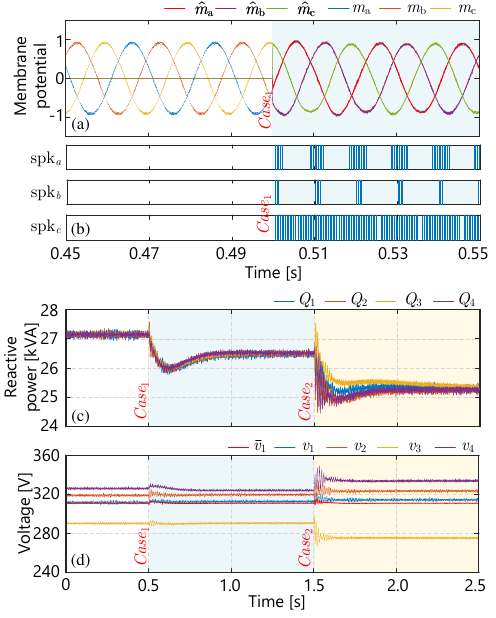}
	\caption{Coordinated control results using the proposed approach in Fig. \ref{fig_1}(c): (a) regression accuracy of the modulation signal, (b) output spikes of SNN, (c) reactive power sharing results, (d) average voltage regulation results.}
    \label{fig_4}
\end{figure}

In Fig. \ref{fig_4}, we firstly illustrate the efficacy of reactive power sharing and average voltage regulation through two test cases: $\mathrm{\texttt{Case}_1}$: a load at Bus \#6 transitions from 0.45 pu to 0.9 pu; $\mathrm{\texttt{Case}_2}$: a line outage occurs between Bus \#9 and \#14. In Fig. \ref{fig_4}(a), the membrane potential $\mathrm{\mathbf{u}_{abc}}$ are obtained from SNN, while the target values are represented as $\mathrm{\mathbf{m}_{abc}}$. Before t = 0.5 s, the system maintains steady state, with $\mathrm{\mathbf{u}_{abc}}$ remaining zero. However, with the onset of $\mathrm{\texttt{Case}_1}$, $\mathrm{\mathbf{u}_{abc}}$ are activated to match $\mathrm{\mathbf{m}_{abc}}$, consequently leading to generation of spikes $\mathrm{\mathbf{spk}_{abc}}$. Since the membrane potential matches the target modulation signal in Fig. \ref{fig_4}(a), the precision behind spike generation for the inverter at Bus \#6 can be formally validated in  Fig. \ref{fig_4}(b). As evident from Fig. \ref{fig_4}(c) and (d), the coordinated control outcomes for both $\mathrm{\texttt{Case}_1}$ and $\mathrm{\texttt{Case}_2}$ are satisfactorily achieved with the average voltage well regulated under several dynamic conditions.

\subsection{Adaptability verification}

\begin{figure}[!t]
    \centering
    \includegraphics[width=0.85\linewidth]{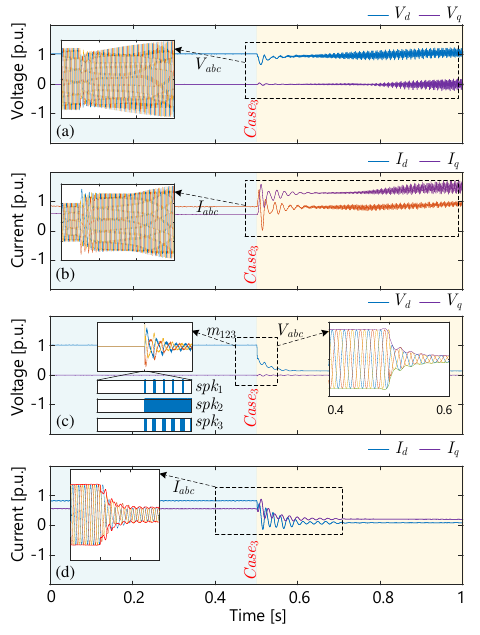}
	\caption{Fault ride through (FRT) compliance: (a) voltage instability without FRT control, (b) overcurrent issues without FRT control, (c) voltage and (d) current adaption.}
    \label{fig_5}
\end{figure}

To illustrate the adaptability of SNN to dynamic conditions, we employ the following cases: $\mathrm{\texttt{Case}_3}$: utility grid voltage sag from 1 pu to 0.1 pu.; $\mathrm{\texttt{Case}_4}$: line outage transpires between Bus \#11 and \#12. Conventionally, handling voltage sags necessitate a mode change between two control mechanisms \cite{liu2021current}. However, in this work, this is carried out in seamless manner that adapts its operation only by sensing declining voltages locally. As depicted in Fig. \ref{fig_5}(a) and (b), the system exhibits instability. However, upon integrating the FRT strategy, voltage and current are adaptively adjusted/safeguarded within the expected limits in response to the extent of the voltage sag, as illustrated in Fig. \ref{fig_5}(c) and (d).

\begin{figure}[!t]
    \centering
	\includegraphics[width=0.85\linewidth]{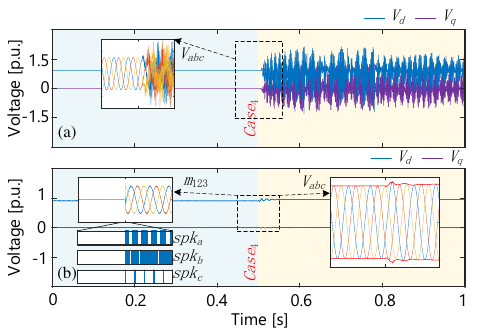}
	\caption{Parameter adaptation: (a) voltage instability without parameters adaption, (b) improved stability by facilitating parameters adaptation using the proposed approach.}
    \label{fig_6}
\end{figure}

Control parameters play a crucial role in system stability. Tuning a single set of parameters for a global set of scenarios is challenging. In system reconfiguration scenarios such as the line outage in $\mathrm{\texttt{Case}_4}$, the system can become unstable if the parameters are not adjusted properly, as illustrated in Fig. \ref{fig_6}(a). Hence, adaptation of SNN weights using power flow dynamics modifies SNN parameters, resulting in the improved stability of the system, as demonstrated in Fig. \ref{fig_6}(b).

\subsection{Energy Consumption Analysis}

To assess the energy efficiency of SNN, we compare its average online operational energy consumption of 90 events with that of artificial neural networks (ANN) and binary-activated recurrent neural networks (RNN), as presented in Table \ref{Table_1}. It is worth notifying that $\mathrm{P_\mathrm{on}}$ and $\mathrm{P_\mathrm{off}}$ denote the power consumption during event-on and event-off states, respectively. Similarly, $\mathrm{N_\mathrm{on}}$ and $\mathrm{N_\mathrm{off}}$ represent the count of active neurons in the event-on and event-off states, respectively with $\mathrm{E_\mathrm{data}}$ denoting the energy consumption of each active neuron. Even though the event-driven operation of binary RNN and SNN are quite similar, they basically differ in their off-time power consumption (in the absence of spikes), where $\mathrm{P_\mathrm{off}}$ is 0 for SNN as compared to 6.06 mW for binary RNN. However under the presence of spikes, $\mathrm{P_\mathrm{on}}$ of binary RNN and SNN are quite similar. As opposed to binary spikes based data processed asynchronously in SNN and binary RNN, ANN operates synchronously to process floating point numbers based data, resulting in considerably higher energy consumption due to ADC stages for numeric values. Consequently, the energy efficiency advantage of SNN over ANN and binary RNN is quite evident.\par

\begin{table}[!t]
\setlength{\tabcolsep}{4pt}
	\centering
	\caption{Comparison of Computational Power}
	\label{Table_1}
    \vspace{6 pt}
	\begin{tabular}{cccccc}
        \hline\hline\noalign{\smallskip}
        & $\mathrm{P_\mathrm{off}}$ [mW] & $\mathrm{P_\mathrm{on}}$ [mW] & $\mathrm{N_\mathrm{on}}$ & $\mathrm{N_\mathrm{off}}$ & $\mathrm{E_\mathrm{data}}$ [pJ]\\
        \noalign{\smallskip}\hline\noalign{\smallskip}
        \texttt{SNN}&0& 8.08& 342.4&0&23.6\\
        Binary-\texttt{RNN}&6.06& 8.50& 360.0&256.6&23.6\\
        \texttt{ANN}&60.80& 60.80& 521&521&116.7\\
        \noalign{\smallskip}\hline\hline
	\end{tabular}
\end{table}

\begin{figure}[!t]\centering
	\includegraphics[width=\linewidth]{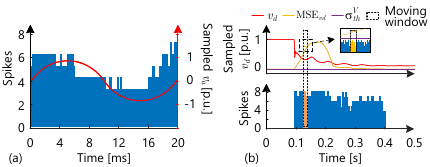}
	\caption{Spikes of hidden layer I in SNN deployed in inverter at Bus \#6: (a) spikes generation alongside the input signal $\mathrm{v_a}$, (b) spikes generation alongside the moving window.}
    \label{fig_7}
\end{figure}

In Fig. \ref{fig_7}(a), the accuracy of asynchronous event-driven sampling is demonstrated over the moving window based on the alignment of input and hidden spikes of SNN with respect to the sinusoidal signal $\mathrm{v_a[m]}$. Since the inputs are sinusoidal signals, the spikes collected in the hidden layer of the SNN are aligned to the input signal as in Fig. \ref{fig_7}(a). In Fig. \ref{fig_7}(b), when the MSE in the moving window of $\mathrm{v_d}$ from (1) is larger than threshold $\mathrm{\sigma_{th}^{V}}$, spikes are continuously generated as per the condition outlined in Algorithm I. 

\subsection{Experimental Results}

\begin{figure}
    \centering
    \includegraphics[width=\linewidth]{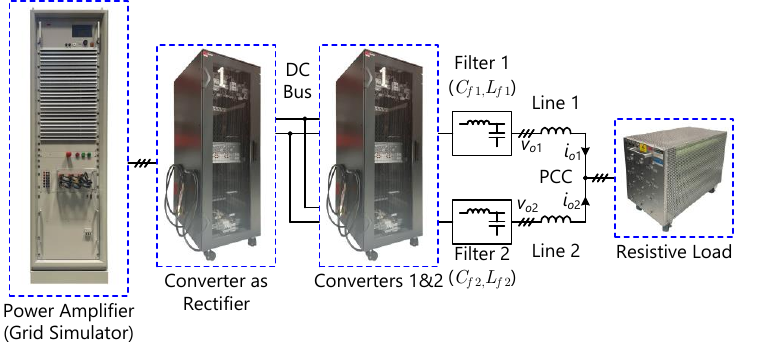}
    \caption{Experimental setup comprising of two three-phase DC–AC converters connected to a resistive load -- the programmed \text{SNN$\mathrm{_1}$} and \text{SNN$\mathrm{_2}$} are deployed into their respective controllers.}
    \label{fig_8}
\end{figure}

The performance of the proposed approach is validated experimentally in the setup shown in Fig. \ref{fig_8}, which illustrates a two-bus islanded testbed. Two 7.5 kVA DC–AC converters (designated as Converters 1 and 2) operating with a reference bus voltage of 110 V are connected to each other via tie-lines to supply a resistive load. The load resistance is 115 $\mathrm{\Omega}$. Since the spikes are generated inside the controller and can not be directly sensed, we recorded the spikes-based data using the Imperix Cockpit and re-plotted it in Fig. 9(d) and 9(h). The designed SNN is given by: \{$\mathrm{\mathbf{Layer}_{in,6\times 1}}$, $\mathrm{\mathbf{Layer}_{hidden,256\times 2}}$, $\mathrm{\mathbf{{Layer}}_{out,3\times 1}}$\}.

\begin{figure}
    \centering
    \includegraphics[width=\linewidth]{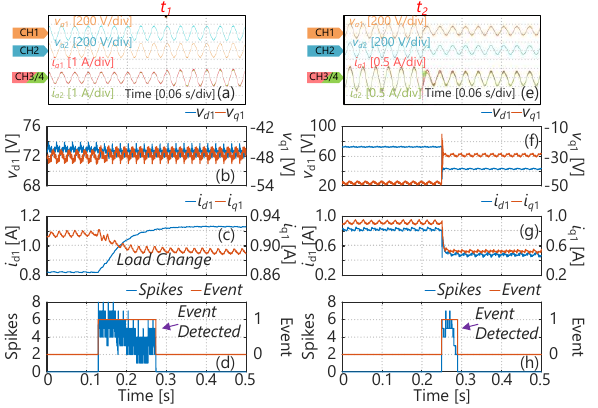}
    \caption{Experimental results of load change and voltage sag using the proposed approach: (a) \& (e) output voltages and currents of Converter 1 and 2 in phase A, (b) \& (f) $\mathrm{v_d}$ and $\mathrm{v_q}$ of converter 1, (c) \& (g) $\mathrm{i_d}$ and $\mathrm{i_q}$ of converter 1, (d) \& (h) Event detection and spikes of the first 8 neurons in the hidden layer are presented.}
    \label{fig_9}
\end{figure}

The results are shown in Fig. \ref{fig_9}. In Fig. \ref{fig_9}(a)-(d), we firstly present the results of load change. In Fig. \ref{fig_9}(a), the load shifts from 115 $\mathrm{\Omega}$ to 75 $\mathrm{\Omega}$ at $\mathrm{t_1}$. The output currents of converters $\mathrm{i_{oa1}}$ and $\mathrm{i_{oa2}}$ increase and are controlled to match. Due to the equivalent reactive load of the filter capacitor and line inductance, $\mathrm{v_{q1}}$ is non-zero in Fig. \ref{fig_9}(b). Considering the load in series with the line inductance and LC filter, resistive load changes cause both active and reactive power dynamics. Consequently, both $\mathrm{i_{d1}}$ and $\mathrm{i_{q1}}$ change, as illustrated in Fig. \ref{fig_9}(c). The detection of $\mathrm{i_{d1}}$, $\mathrm{i_{q1}}$, $\mathrm{v_{d1}}$, and $\mathrm{v_{q1}}$ dynamics triggers events as per (\ref{eq_event}). Consequently, the spikes originating from the event-driven sampling are hereby generated. For a clear presentation of spikes, only the first 8 neurons are presented in the hidden layer. Finally, when the event is inactive, the SNN enters into an idle state.

In Fig. \ref{fig_9}(e)-(h), we present the results of a voltage sag scenario. Converter 2's output voltage is deliberately lowered to test the adaptive response of converter 1. In Fig. \ref{fig_9}(e), converter 1's output voltage $\mathrm{v_{a1}}$ adeptly tracks $\mathrm{v_{a2}}$ at $\mathrm{t_2}$. This voltage matching results in controlled lower values for the two output currents, $\mathrm{i_{oa1}}$ and $\mathrm{i_{oa2}}$, without excessive overcurrent. The values of $\mathrm{v_{d1}}$ and $\mathrm{v_{q1}}$ in Fig. \ref{fig_9}(f) shift closer to zero. Due to the voltage sag, both $\mathrm{i_{d1}}$ and $\mathrm{i_{q1}}$ decrease, as illustrated in Fig. \ref{fig_9}(g). The detection of dynamics in $\mathrm{i_{d1}}$, $\mathrm{i_{q1}}$, $\mathrm{v_{d1}}$, and $\mathrm{v_{q1}}$ triggers the events, as shown in Fig. \ref{fig_9}(h). The time scale of spikes aligns coherently with the detected event.


\section{Conclusions and Future scope of work}
This paper leverages neuromorphic computing principles using a SNN at each grid-edge power electronic converter that enables them to comprehend grid disturbances by observing local power dynamics across the tie-lines in an asynchronous, event-driven manner and provide online adaptive response to it accordingly. As opposed to the state-of-the-art grid-edge technologies, it offers the following advantages:
\begin{enumerate}
\item It dismisses coordination via cyber layer, thereby preventing exogenous path arrivals for cyber attackers and further security concerns.
\item It dismisses the cyber unreliability issues and its impact on power electronic dominated grids.
\item It is an energy efficient distributed learning and grid-edge intelligence tool, and it doesn't suffer from the system and transmission efficiency issues beyond electrically isolated units.
\end{enumerate}
As a future scope of work, we plan to explore the effectiveness of its powerful online learning capability against some of the unseen grid conditions and establish system resiliency without communications as an overarching layer. 

\ifCLASSOPTIONcaptionsoff
  \newpage
\fi

\bibliography{Reference.bib}

\begin{thebibliography}{10}
\providecommand{\url}[1]{#1}
\csname url@samestyle\endcsname
\providecommand{\newblock}{\relax}
\providecommand{\bibinfo}[2]{#2}
\providecommand{\BIBentrySTDinterwordspacing}{\spaceskip=0pt\relax}
\providecommand{\BIBentryALTinterwordstretchfactor}{4}
\providecommand{\BIBentryALTinterwordspacing}{\spaceskip=\fontdimen2\font plus
\BIBentryALTinterwordstretchfactor\fontdimen3\font minus \fontdimen4\font\relax}
\providecommand{\BIBforeignlanguage}[2]{{%
\expandafter\ifx\csname l@#1\endcsname\relax
\typeout{** WARNING: IEEEtran.bst: No hyphenation pattern has been}%
\typeout{** loaded for the language `#1'. Using the pattern for}%
\typeout{** the default language instead.}%
\else
\language=\csname l@#1\endcsname
\fi
#2}}
\providecommand{\BIBdecl}{\relax}
\BIBdecl

\bibitem{sahoo}
S.~Sahoo and F.~Blaabjerg, ``{A Model-Free Predictive Controller for Networked Microgrids with Random Communication Delays},'' in \emph{2021 IEEE Applied Power Electronics Conference and Exposition (APEC)}, 2021, pp. 2667--2672.

\bibitem{cybbook}
S.~Sahoo, T.~Dragičević, and F.~Blaabjerg, ``{Cyber Security in Control of Grid-Tied Power Electronic Converters—Challenges and Vulnerabilities},'' \emph{IEEE J. Emerging Sel. Top. Power Electron.}, vol.~9, no.~5, pp. 5326--5340, 2021.

\bibitem{ye}
J.~Ye \emph{et~al.}, ``{A Review of Cyber--Physical Security for Photovoltaic Systems},'' \emph{IEEE J. Emerging Sel. Top. Power Electron.}, vol.~10, no.~4, pp. 4879--4901, 2021.

\bibitem{sahoo2022cyber}
S.~Sahoo, F.~Blaabjerg, and T.~Dragičević, \emph{{Cyber Security for Microgrids}}.\hskip 1em plus 0.5em minus 0.4em\relax Institution of Engineering and Technology, 2022.

\bibitem{yubo}
Y.~Song, S.~Sahoo, Y.~Yang, and F.~Blaabjerg, ``{Stability Constraints on Reliability-Oriented Control of AC Microgrids--Theoretical Margin and Solutions},'' \emph{IEEE Trans. Power Electron.}, vol.~38, no.~8, pp. 9459--9468, 2023.

\bibitem{7500071}
J.~W. Simpson-Porco, F.~Dörfler, and F.~Bullo, ``{Voltage Stabilization in Microgrids via Quadratic Droop Control},'' \emph{IEEE Trans. Autom. Control}, vol.~62, no.~3, pp. 1239--1253, 2017.

\bibitem{PowerTalk}
M.~Angjelichinoski, C.~Stefanovic, P.~Popovski, H.~Liu, P.~C. Loh, and F.~Blaabjerg, ``{Power Talk: How to Modulate Data over a DC Micro Grid Bus Using Power Electronics},'' in \emph{Proc. 2015 IEEE Global Communications Conference (GLOBECOM)}, 2015, pp. 1--7.

\bibitem{marco}
M.~Liserre, H.~Beiranvand, Y.~Leng, R.~Zhu, and P.~A. Hoeher, ``{Overview of Talkative Power Conversion Technologies},'' \emph{IEEE Open J. Power Electron.}, vol.~4, pp. 67--80, 2023.

\bibitem{sahoo2024nsc}
X.~Diao, Y.~Song, S.~Sahoo, and Y.~Li, ``{Neuromorphic Event-Driven Semantic Communication in Microgrids},'' \emph{IEEE Trans. Smart Grid}, 2024, early access.

\bibitem{sahoo2024spike}
S.~Sahoo, ``{Spike Talk: Genesis and Neural Coding Scheme Translations},'' \emph{arXiv preprint arXiv:2408.00773}, 2024.

\bibitem{kirti1}
K.~Gupta, S.~Sahoo, and B.~K. Panigrahi, ``{A Monolithic Cybersecurity Architecture for Power Electronic Systems},'' \emph{IEEE Trans. Smart Grid}, vol.~15, no.~4, pp. 4217--4227, 2024.

\bibitem{kirti2}
K.~{Gupta}, S.~Sahoo, and B.~K. Panigrahi, ``{Delay-Aware Semantic Sampling in Power Electronic Systems},'' \emph{IEEE Trans. Smart Grid}, vol.~15, no.~4, pp. 4038--4049, 2024.

\bibitem{EventDriven}
J.~Chen, N.~Skatchkovsky, and O.~Simeone, ``{Neuromorphic Wireless Cognition: Event-Driven Semantic Communications for Remote Inference},'' \emph{IEEE Trans. Cognit. Commun. Networking}, vol.~9, no.~2, pp. 252--265, 2023.

\bibitem{loihi}
M.~Davies, A.~Wild, G.~Orchard, Y.~Sandamirskaya, G.~A.~F. Guerra, P.~Joshi, P.~Plank, and S.~R. Risbud, ``{Advancing Neuromorphic Computing With Loihi: A Survey of Results and Outlook},'' \emph{Proc. IEEE}, vol. 109, no.~5, pp. 911--934, 2021.

\bibitem{skatchkovsky2021spiking}
N.~Skatchkovsky, H.~Jang, and O.~Simeone, ``{Spiking Neural Networks—Part II: Detecting Spatio-Temporal Patterns},'' \emph{IEEE Commun. Lett.}, vol.~25, no.~6, pp. 1741--1745, 2021.

\bibitem{liu2021current}
T.~Liu, X.~Wang, F.~Liu, K.~Xin, and Y.~Liu, ``{A Current Limiting Method for Single-loop Voltage-magnitude Controlled Grid-forming Converters During Symmetrical Faults},'' \emph{IEEE Trans. Power Electron.}, vol.~37, no.~4, pp. 4751--4763, 2021.

\bibitem{para}
S.~K.~M. Kodsi and C.~A. Cañizares, ``{Modeling and Simulation of IEEE 14-Bus System with FACTS Controllers},'' in \emph{University of Waterloo, Canada, Tech. Rep.}, 2003.

\end{thebibliography}
\bibliographystyle{IEEEtran}

\end{document}